%% file: main-r3camera.tex
\documentclass[10pt,journal,letterpaper]{IEEEtran}

\usepackage{cite}
\usepackage{amsmath,amssymb,amsfonts}
\usepackage{amsthm}
\usepackage{algorithmic}
\usepackage{graphicx}
\usepackage{textcomp}
\usepackage{xcolor}

\usepackage{url}
\usepackage{bm} % For bold math vectors

\usepackage{tikz}
\usetikzlibrary{arrows.meta, decorations.pathmorphing, patterns}

\theoremstyle{plain}

\theoremstyle{definition}
\newtheorem{remark}{Remark}

\title{On the Tail Transition of First Arrival Position Channels: From Cauchy to Exponential Decay}

\author{
    Yen-Chi~Lee,~\IEEEmembership{Member,~IEEE}~
    \thanks{This work was supported by the National Science and Technology Council of Taiwan (NSTC 113-2115-M-008-013-MY3). (Corresponding author: Yen-Chi Lee.)}
    \thanks{Y.-C. Lee is with the Department of Mathematics, National Central University, Taoyuan, Taiwan (e-mail: \texttt{yclee@math.ncu.edu.tw}).}
}

\usepackage[
    colorlinks=true,        % Use colored links instead of boxes
    linkcolor=blue,         % Internal links (sections, equations, theorems)
    citecolor=blue,         % Citation links
    urlcolor=blue,          % URL links
    bookmarks=true,         % Enable PDF bookmarks
    pdfborder={0 0 0}       % Remove link borders
]{hyperref}

\begin{document}

\maketitle

% ==============================================================================
% ABSTRACT & KEYWORDS
% ==============================================================================
\begin{abstract}
    While the zero-drift first arrival position (FAP) channel exhibits a Cauchy-distributed lateral displacement, nonzero drift in practical systems introduces advective transport that regularizes this singular limit. This letter characterizes the drift-induced transition of FAP distribution from heavy-tailed algebraic regime to exponential regularization. By asymptotically examining the exact FAP density, we identify a \textit{characteristic propagation distance (CPD)} that serves as the fundamental boundary separating diffusion-dominated and drift-dominated regimes. Numerical experiments demonstrate that in low-drift environments, variance-matched Gaussian approximations severely underestimate the true communication potential, whereas the zero-drift Cauchy law provides a robust, physically grounded performance baseline.
\end{abstract}

\begin{IEEEkeywords}
    Molecular communications (MC), first arrival position (FAP), Cauchy distribution, heavy-tailed noise, achievable rate.
\end{IEEEkeywords}

% ==============================================================================
% SECTION I: INTRODUCTION
% ==============================================================================
\section{Introduction}\label{sec:intro}

\IEEEPARstart{M}{olecular} communication via diffusion (MCvD) relies on the stochastic motion of information molecules to convey messages~\cite{Akyildiz2008,Nakano2013,Farsad2016}. While most existing studies focus on \textit{first hitting time (FHT)}~\cite{Kadloor2012,Srinivas2012,Yilmaz2014,Lee2026}, the spatial dimension---namely, the \textit{first arrival position (FAP)}~\cite{Lee2016,Pandey2018,Lee2024_TCOM}---offers an alternative information type. Pure FAP channels can be fundamentally viewed as event-driven systems \cite{Hsieh2012,Hsieh2013}, as the receiver performs decisions solely at the instants of particle arrival (i.e., the decision times) based on spatial impact coordinates. 

Under the framework of asynchronous information embedding \cite{Hsieh2012,Hsieh2013}, the timing of these events---namely the FHT---serves as an orthogonal signaling scheme for such event-driven systems. This implies that additional information can be embedded directly into the arrival times without influencing the original spatial-based decision outcomes, inherently allowing for a higher information rate through joint spatio-temporal modulation.

Beyond single-link communication, FAP can be viewed as a fundamental \textit{spatial impulse response} that governs the design of spatial signaling in molecular MIMO~\cite{Koo2016} and index modulation~\cite{Gursoy2019} systems.
Furthermore, recent microfluidic surveys~\cite{Hamidovic2024} underscore the impact of spatial effects, where FAP statistics directly govern cross-talk between adjacent sensors. Since these interactions limit feasible receiver density, a precise characterization of FAP distribution becomes indispensable for robust system design and interference management.

In previous work~\cite{Lee2024_ICC}, the authors established that zero-drift FAP communication channels exhibit a heavy-tailed Cauchy noise law. However, this singular limit is often regularized in practical environments by nonzero drift, which introduces advective transport \cite{Jamali2019}. This regularization mechanism is rooted in the one-dimensional (1-D) FHT characteristics: under drift, the FHT is governed by an Inverse Gaussian (IG) distribution that peaks significantly earlier than the zero-drift L\'{e}vy distribution \cite{Kadloor2012,Srinivas2012}. This temporal advancement ensures particles reach the receiver more rapidly, effectively \textit{truncating the time available for transverse random walks to develop}. Consequently, lateral displacement is confined, suppressing the Cauchy heavy-tailed phenomenon and regularizing the tail into an exponential decay.

Since practical systems typically operate in intermediate regimes where drift and diffusion coexist \cite{Jamali2019}, accounting for this tail transition is vital to avoid misleading design conclusions. In particular, a heuristic selection of a variance-matched Gaussian model may lead to overly conservative performance estimates in low-drift regimes, whereas a pure Cauchy baseline fails to account for the exponential regularization as advective transport increases \cite{Lee2024_ICC,Jamali2019}. Such discrepancies underscore the need for a unified framework that bridges these two physical extremes.

The aim of this letter is to characterize the drift-induced transition of the FAP channel from heavy-tailed algebraic regime to exponential regularization. Through asymptotic analysis of the exact FAP distribution, we identify a \textit{characteristic propagation distance (CPD)}, which separates diffusion-dominated and drift-dominated regimes. Numerical experiments further establish that the zero-drift Cauchy law serves as a physical baseline for achievable-rate analysis in low-drift environments, where Gaussian approximations fail to capture the true operational limits of the channel.

The remainder of this letter is organized as follows. Section~\ref{sec:system} introduces the system model and the exact FAP distributions. Section~\ref{sec:tail} presents an asymptotic analysis of the tail behavior and identifies the CPD. Section~\ref{sec:numerical} investigates the implications of this transition for achievable rate and spatial interference. Finally, Section~\ref{sec:conclude} concludes the letter.

% ==============================================================================
% SECTION II: SYSTEM MODEL AND EXACT DISTRIBUTION
% ==============================================================================
\section{System Model and Exact Distribution}\label{sec:system}

We consider an MCvD system in $\mathbb{R}^d$, with the $d=2$ case illustrated in Fig.~\ref{fig:channel_model}. The transmitter (Tx) releases particles at $\mathbf{X} \in \mathbb{R}^{d-1}$ on a hyperplane. Each particle undergoes Brownian motion with diffusion coefficient $D$ (with variance $\sigma^2 = 2D$) and a constant drift velocity $\mathbf{v}$ perpendicular to the receiving hyperplane at distance $\lambda > 0$. Let $v = \|\mathbf{v}\|$ denote the drift speed.

Information particle arrivals at the receiver (Rx) are characterized by transverse coordinates $\mathbf{Y} = \mathbf{X} + \mathbf{N}$, where $\mathbf{N} \in \mathbb{R}^{d-1}$ is the random lateral displacement. Let $\mathbf{n}$ denote a realization of $\mathbf{N}$. The \emph{impact position} is measured relative to the point directly facing the release position; thus, the origin $\mathbf{N} = \mathbf{0}$ represents central impacts. For $d=2$, we simplify to the scalar form $Y = X + N$.

For clarity, the main notations used throughout this letter are summarized in Table~\ref{tab:notations}.

\input{TABLE1}

\subsection{The Zero-drift Baseline: Cauchy Distribution}
\label{subsec:cauchy_baseline}

The noise distribution is governed by the drift velocity $\mathbf{v}$. In the zero-drift limit ($v\to 0$), the FAP noise reduces to a multivariate Cauchy distribution~\cite{Lee2024_ICC}. For the $d=2$ case (scalar noise $N$), this drift-free baseline is given by \cite{Lee2024_ICC},
\begin{equation}\label{eq:cauchy_limit}
    \lim_{v\to 0^+} f_N(n\mid v) = \frac{\lambda}{\pi\left(n^2+\lambda^2\right)}.
\end{equation}
Equation~\eqref{eq:cauchy_limit} serves as a physical reference point: in the absence of external flow, the channel exhibits an intrinsically (algebraic) heavy-tailed Cauchy law.

\subsection{Non-Zero Drift Case: Exact FAP Distribution}
\label{subsec:general_drift}

For $v>0$ (pointing directly to the Tx and assisting the transmission, as shown in Fig.~\ref{fig:channel_model}), the exact probability density function (PDF) of the random lateral displacement for the $d=2$ case is available in~\cite{Lee2024_TCOM} and can be written as
\begin{equation}\label{eq:exact_pdf}
    \begin{aligned}
        f_N(n \mid v) &= \frac{v\lambda}{\pi \sigma^2 \sqrt{n^2 + \lambda^2}} \exp\!\left(\frac{v\lambda}{\sigma^2}\right) K_1\!\left( \frac{v}{\sigma^2}\sqrt{n^2+\lambda^2} \right).
    \end{aligned}
\end{equation}
Here, $K_1(\cdot)$ denotes the modified Bessel function of the second kind. Although \eqref{eq:exact_pdf} is exact, its complicated Bessel function form obscures the physical intuition behind tail regularization under drift. We therefore develop an asymptotic analysis in the next section to expose the underlying structure.

% --- FIGURE 1: MIRRORED SYSTEM MODEL ---
\begin{figure}[!t]
    \centering
    \resizebox{0.45\textwidth}{!}{%
        \begin{tikzpicture}[
            >=Latex,
            node distance=2cm,
            font=\sffamily\small
        ]
            \def\TxX{0}      
            \def\RxX{5}      
            \def\Ymax{2.5}
            \def\Ymin{-2.5}

            % --- Receiver (Rx) ---
            \draw[blue!80!black, very thick] (\RxX, \Ymin) -- (\RxX, \Ymax) node[above] {Rx};
            \fill[pattern=north east lines, pattern color=blue!30] (\RxX, \Ymin) rectangle (\RxX+0.3, \Ymax);

            % --- Transmitter (Tx) ---
            \draw[red!80!black, very thick, dashed] (\TxX, \Ymin) -- (\TxX, \Ymax) node[above] {Tx};

            % --- Information Source ---
            \node[draw, fill=yellow!90!orange, minimum width=2cm, minimum height=1cm, align=center] (source) at (\TxX-2.5, 0) {Information\\Source};
            \draw[->, thick] (source) -- (\TxX, 0);

            % --- Information Sink ---
            \node[draw, fill=yellow!90!orange, minimum width=2cm, minimum height=1cm, align=center] (sink) at (\RxX+2.5, 0) {Information\\Sink};
            \draw[<-, thick] (sink) -- (\RxX, 0);

            % --- Particle Path ---
            \fill[red] (\TxX, 0.5) circle (3pt); 
            \fill[blue] (\RxX, 1.2) circle (3pt); 
            
            \draw[red, thick, decorate, decoration={random steps,segment length=3pt,amplitude=4pt}] 
                (\TxX, 0.5) -- (1.5, 0.8) -- (3.0, 0.2) -- (4.0, 1.5) -- (\RxX, 1.2);
                
            \node[red!80!black, font=\footnotesize] at (2.5, 1.8) {Trajectory of Information Particle};

            % --- Drift Velocity Arrow ---
            \draw[->, thick, color=red!60!black] (1.5, -0.5) -- (3.5, -0.5) 
                node[midway, below, font=\footnotesize, align=center] {Drift Velocity $\mathbf{v}$ \\ (Assisting the transmission)};

            % --- Lambda Dimension ---
            \draw[<->, thick] (\TxX, \Ymin-0.3) -- (\RxX, \Ymin-0.3) node[midway, below] {$\lambda$};

            % --- Bottom Dashed Lines ---
            \draw[gray, thin, dashed] (\RxX, \Ymin) -- (\RxX, \Ymin-0.5);
            \draw[gray, thin, dashed] (\TxX, \Ymin) -- (\TxX, \Ymin-0.5);

        \end{tikzpicture}%
    }
    \caption{
        A conceptual figure of FAP channel model featuring hyperplane-shaped Tx and Rx pair. The information particle undergoes Brownian motion with diffusion coefficient $D$ and drift velocity $\mathbf{v}$. The longitudinal transmission distance is $\lambda$. The Rx observes and records the arrival point (i.e., impact position) on the transverse hyperplane.
    }
    \label{fig:channel_model}
\end{figure}

% ==============================================================================
% SECTION III: THE ANATOMY OF THE TAIL
% ==============================================================================
\section{The Anatomy of the Tail Transition: Asymptotic Analysis}\label{sec:tail}

The tail of the noise distribution is governed by the argument of the modified Bessel function in \eqref{eq:exact_pdf}. For brevity, we analyze the 2-D ($d=2$) case. We introduce the radial propagation distance 
\( 
    r = r(n) := \sqrt{n^2+\lambda^2}
\) 
and the dimensionless parameter 
\( 
    z = z(n) := \frac{r(n)}{r_c},
\)
where $r_c = \sigma^2/v$ is the CPD representing the spatial scale at which drift and diffusion effects are balanced.

\subsection{Regime 1: The Cauchy Core (Diffusion-Dominated)}
\label{subsec:regime1}

We first consider the small-$z$ regime, defined by $z(n)\ll 1$, which corresponds to weak drift or small radial distance $r(n)$. In this regime, the modified Bessel function admits the small-argument asymptotic expansion $K_1(z)\sim 1/z$ (see Appendix~\ref{app:bessel_small}). Substituting this approximation into the exact expression \eqref{eq:exact_pdf} gives
\begin{equation}
    \begin{split}
        f_N(n \mid v) &\sim \frac{v\lambda}{\pi \sigma^2 r(n)} \left( \frac{\sigma^2}{v r(n)} \right), \qquad z(n)\to 0, \\
                      &\sim \frac{\lambda}{\pi (n^2+\lambda^2)}, \qquad\qquad\quad~z(n)\to 0,
    \end{split}
\end{equation}
where we also used $\exp(v\lambda/\sigma^2)\to 1$ as $v\to 0$. Therefore, in the small-$z$ limit, the noise distribution reduces to a Cauchy-type law~\cite{Lee2024_ICC}, characterized by an algebraic heavy tail of order $O(n^{-2})$. Physically, this regime represents a diffusion-dominated core, in which random thermal motion overwhelms the effect of drift and the channel behavior is well approximated by the zero-drift baseline described in Section~\ref{subsec:cauchy_baseline}.

\subsection{Regime 2: The Drift-Regularized Tail (Drift-Dominated)}
\label{subsec:regime2}

We next consider the large-$z$ regime, defined by $z(n)\gg 1$, which corresponds to large radial distances $r(n)$ or sufficiently strong drift. In this regime, the modified Bessel function admits the large-argument asymptotic expansion $K_1(z)\sim \sqrt{\pi/(2z)}\,e^{-z}$ as $z\to\infty$ (see Appendix~\ref{app:bessel_large}). Substituting this into the exact expression \eqref{eq:exact_pdf} and applying $z(n)=(v/\sigma^2)r(n)$, the prefactor simplifies to yield
\begin{equation}\label{eq:regime2_tail}
    f_N(n\mid v) \sim \frac{\lambda}{\sigma}\sqrt{\frac{v}{2\pi}}\, r(n)^{-3/2} \exp\!\left( \frac{v}{\sigma^2}\bigl(\lambda-r(n)\bigr) \right).
\end{equation}
Recalling $r(n)=\sqrt{n^2+\lambda^2}$, \eqref{eq:regime2_tail} provides the explicit algebraic-exponential decay in terms of the lateral coordinate $n$. 

Furthermore, for large transverse displacements $|n|\gg \lambda$, we have $r(n) \sim |n|$. The asymptotic form in \eqref{eq:regime2_tail} thus reduces to show that the tail is exponentially regularized as
\begin{equation} 
    f_N(n\mid v) = \mathcal{O}\!\left(|n|^{-3/2}\exp\!\left(-\frac{v}{\sigma^2}|n|\right)\right), \qquad |n|\to\infty.
\end{equation}
Consequently, $\mathbb{E}|N|^p<\infty$ for all $p>0$ whenever $v>0$.

Taken together, the small-$z$ and large-$z$ regimes reveal the full anatomy of the noise distribution: diffusion gives rise to a Cauchy-like core near the origin, while drift progressively regularizes the tail by exponentially suppressing large transverse displacements.

\subsection{The Critical Length Scale}
\label{subsec:critical_scale}

The crossover between the diffusion-dominated Cauchy core and the drift-regularized tail is governed by the dimensionless parameter $z(n)=r(n)/r_c$. In particular, the transition occurs when $z(n)$ is of order unity, i.e.,
\begin{equation}\label{eq:critical_scale}
    z(n) \sim \mathcal{O}(1) \quad \Leftrightarrow \quad r(n) \sim r_c .
\end{equation}
Physically, $r_c$ acts as a cutoff length separating two noise regimes: when $r(n)\ll r_c$ (equivalently, $z(n)\ll 1$), the channel is well approximated by the Cauchy-like diffusion core; when $r(n)\gg r_c$ (equivalently, $z(n)\gg 1$), the tail is exponentially regularized by drift. For large transverse displacements $|n|\gg \lambda$, since $r(n)\sim |n|$, this cutoff can be interpreted directly in terms of the lateral coordinate.

\subsection{Extension to 3-D Systems}
\label{subsec:3d_extension}

Although the above derivations focus on a 2-D system with a 1-D receiving line for clarity, the tail anatomy is not dimension-specific. The same mechanism extends naturally to 3-D systems with a 2-D receiving plane, which are of practical relevance to planar receiver architectures (see \cite[Fig.~1]{Gursoy2019}).

In a 3-D setting, the noise becomes a vector $\mathbf{N}\in\mathbb{R}^2$, and the exact FAP density depends on the Euclidean propagation distance
\(
    r(\mathbf{n}) := \sqrt{\|\mathbf{n}\|^2+\lambda^2}
\)
with tail behavior governed by an exponential damping factor of the form $\exp\bigl(-v\,r(\mathbf{n})/\sigma^2\bigr)$~\cite{Lee2024_TCOM}. Here we assume that the drift is pointing directly to the Tx (normal to the Rx plane), and $v>0$ denotes the magnitude of the drift vector. Applying the same asymptotic framework yields two analogous regimes:
\begin{itemize}
    \item \textbf{Core ($r(\mathbf{n})\ll r_c$):} The distribution converges to a bivariate Cauchy law, exhibiting an algebraic decay $\propto \|\mathbf{n}\|^{-3}$ and providing two spatial degrees of freedom (see \cite{Lee2024_ICC}).
    \item \textbf{Tail ($r(\mathbf{n})\gg r_c$):} The distribution undergoes exponential regularization, ensuring the existence of finite moments (see \cite[Section~IV]{Lee2024_TCOM}).
\end{itemize}
Importantly, the same characteristic distance $r_c=\sigma^2/v$ governs the crossover between heavy-tailed and light-tailed behavior, independent of the dimensionality of the receiving geometry.

% ==============================================================================
% SECTION IV: NUMERICAL RESULTS
% ==============================================================================
\section{Numerical Implications of the Tail Transition}\label{sec:numerical}

This section evaluates the numerical implications of the tail transition for FAP-based MCvD system performance. We focus on two key metrics of direct engineering relevance: (i) the achievable information rate under peak-amplitude constraints, and (ii) the spatial interference probability, which characterizes cross-talk in molecular MIMO systems (see \cite{Koo2016,Gursoy2019}).

\subsection{Numerical Setup}

Numerical evaluations are performed with transmission distance $\lambda=10$, peak-amplitude constraint $a=200$, and diffusion coefficient $D=100$ (normalized units). The achievable information rate is computed as the mutual information $I(X;Y)$ assuming a uniform input distribution $X\sim\mathcal{U}[-a,a]$, which represents a standard and tractable achievable-rate benchmark under peak constraints.

For comparison, we also include a Gaussian approximation based on a variance-matched \textit{additive white Gaussian noise (AWGN)} model. Specifically, the corresponding achievable rate is given by
\begin{equation}
    R_{\mathrm{Gauss}} = \frac{1}{2}\ln\!\left(1+\frac{P_X}{\sigma_N^2(v)}\right),
\end{equation}
where $\sigma_N^2(v)$ denotes the noise variance computed numerically from the exact distribution in~\eqref{eq:exact_pdf}, and $P_X=a^2/3$ is the average power of the uniform input. This construction ensures that the Gaussian model is informed of the true second-order noise statistics, thereby isolating the effect of heavy-tailed behavior beyond variance alone.

For clarity, we focus on the 1-D transverse setting ($d=2$), which most transparently exhibits the tail-transition phenomenon.

\begin{remark}[Numerical Stabilization]
    Since the $n$-dependence of~\eqref{eq:exact_pdf} is entirely captured by the Bessel-kernel term, we evaluate a kernel proportional to $K_1(z(n))/r(n)$, which is numerically normalized so that the resulting density integrates to one. For very small drift velocities, the integration grid is refined near $n=0$ to accurately capture the singular behavior $K_1(z)\sim 1/z$. Throughout, integral-based evaluations are treated as the numerical reference, while particle-based simulations (PBS) are used solely for independent validation and illustration purposes.
\end{remark}

% --- FIGURE 2: AR RESULTS ---
\begin{figure}[t]
    \centering
    \includegraphics[width=0.95\linewidth]{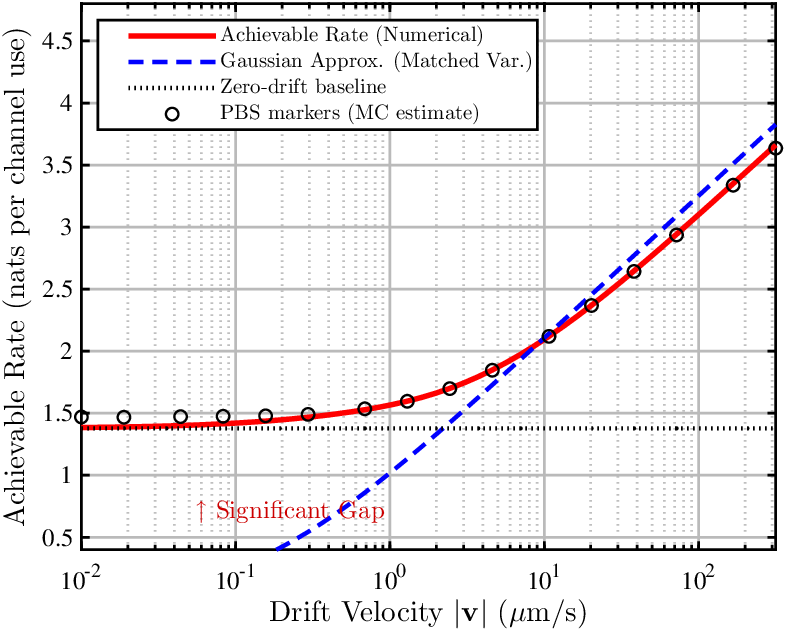}
    \caption{Achievable rate of the FAP channel versus drift velocity $|\mathbf{v}|$ under a uniform input constraint. Solid, dashed, and dotted lines represent numerical results, the matched-variance Gaussian approximation, and the zero-drift baseline, respectively. Markers denote PBS results for validation. Minor PBS overestimation at low drift is expected due to the slow convergence of entropy estimators under heavy-tailed noise.}
    \label{fig:capacity}
\end{figure}
% ----------------------------------

\subsection{Achievable Rate Results}

Fig.~\ref{fig:capacity} illustrates the achievable information rate as a function of the drift velocity. Three curves are shown: the numerically evaluated achievable rate under the uniform input $X \sim \mathcal{U}[-a,a]$ (solid), the variance-matched Gaussian approximation $R_{\mathrm{Gauss}}$ (dashed), and the zero-drift baseline (dotted), all evaluated under the same peak constraint.

Two key observations emerge. First, in the low-drift regime, the achievable rate does not vanish as $v\to 0$; instead, it stabilizes and approaches a finite baseline value. This behavior indicates that reliable information transmission remains possible in diffusion-dominated regimes, even in the absence of drift assistance. Second, the Gaussian approximation significantly underestimates the achievable rate at low drift. As the FAP distribution approaches the Cauchy limit, its variance diverges, causing variance-based Gaussian models to predict vanishing rates, despite the presence of substantial information-carrying capability. This discrepancy highlights the fundamental inadequacy of heuristic Gaussian approximations for heavy-tailed FAP channels.

Collectively, these findings demonstrate that in diffusion-dominated regimes---where $r_c$ exceeds the relevant array spacing or aperture scale---variance-based Gaussian approximations tend to be overly pessimistic, failing to capture the channel's true operational limits.

\begin{remark}[PBS Validation]
    Discrete PBS results serve as an independent validation, though minor deviations at low drift (see Fig.~\ref{fig:capacity}) are expected due to the slow convergence of entropy estimators under heavy-tailed noise. Accuracy improves rapidly as $|\mathbf{v}|$ increases.
    
    %as drift-induced regularization facilitates convergence to the theoretical predictions.
\end{remark}

% --- FIGURE 3: INTERFERENCE RESULTS ---
\begin{figure}[t]
    \centering
    \includegraphics[width=0.89\linewidth]{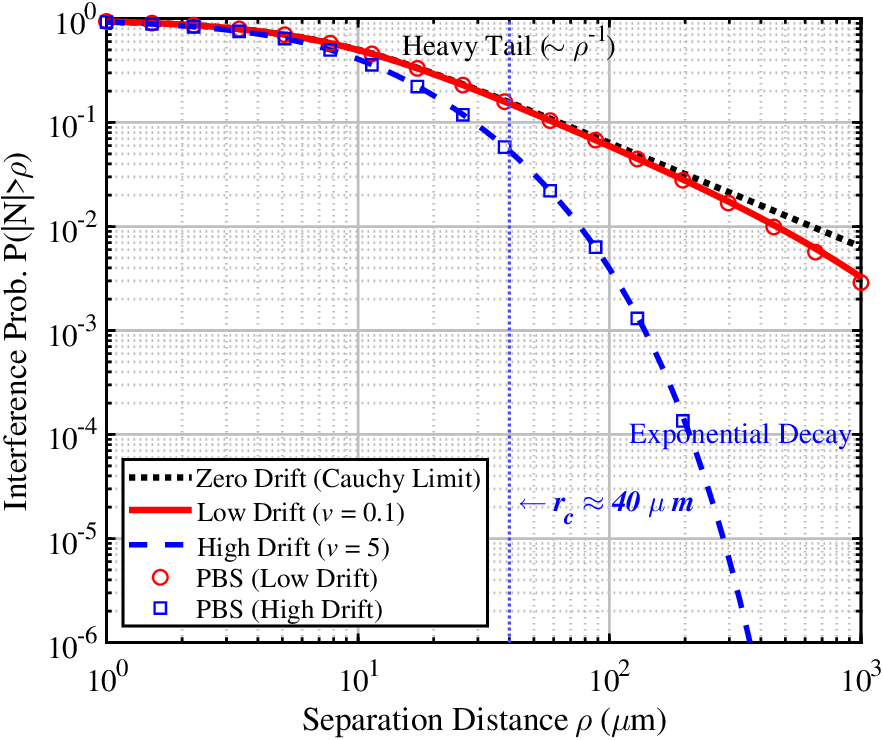}
    \caption{Spatial interference probability $\mathbb{P}(|N|>\rho)$ versus lateral separation distance $\rho$. Theoretical curves are obtained from the exact FAP law, while discrete markers indicate PBS results based on FHT sampling. The characteristic distance $r_c$ delineates the crossover from diffusion-dominated algebraic decay to drift-regularized exponential decay.}
    \label{fig:interference}
\end{figure}
% ----------------------------------

\subsection{Impact on Spatial Interference}

To quantify spatial cross-talk in molecular MIMO systems, we consider the spatial interference probability
\begin{equation}
    P_{\mathrm{int}}(\rho)=\mathbb{P}(|N|>\rho),
\end{equation}
which represents the likelihood that a particle arrives beyond a lateral separation distance $\rho$ from its intended receiver location.

Fig.~\ref{fig:interference} presents numerical results for both zero-drift and nonzero-drift scenarios. In the zero-drift case, the transverse FAP $N$ follows the Cauchy distribution as given in~\eqref{eq:cauchy_limit}. The corresponding interference probability therefore follows directly from integration of the Cauchy density,
\begin{align}
    \begin{split}
        P_{\mathrm{int}}(\rho) &= \mathbb{P}(|N|>\rho) = 2\int_{\rho}^{\infty} \lim_{v\to 0^{+}} f_N(n\mid v)\,\mathrm{d}n \\
                               &= \frac{2}{\pi} \left[ \arctan\!\left(\frac{n}{\lambda}\right) \right]_{\rho}^{\infty} = 1-\frac{2}{\pi}\arctan\!\left(\frac{\rho}{\lambda}\right).
    \end{split}
\end{align}
For large separation distances $\rho\gg\lambda$, this expression admits the asymptotic form
\(
    P_{\mathrm{int}}(\rho) \sim \frac{2\lambda}{\pi\rho},
\)
confirming the algebraic $\rho^{-1}$ decay characteristic of diffusion-dominated transport.

Under nonzero drift, $P_{\mathrm{int}}(\rho)$ is evaluated numerically as
\begin{equation}
    P_{\mathrm{int}}(\rho) = 2\int_{\rho}^{\infty} f_N(n\mid v)\,\mathrm{d}n,
\end{equation}
leveraging the symmetry of the FAP distribution to simplify the integration. The resulting curves reveal a systematic shift in spatial interference behavior as drift strength increases. In drift-dominated regimes, $P_{\mathrm{int}}(\rho)$ decays rapidly once the separation distance exceeds the characteristic scale $r_c = \sigma^2/v$ (see Fig.~\ref{fig:interference}), which provides a physically grounded guideline for optimizing receiver spacing in dense arrays. Conversely, when $r_c$ exceeds relevant system dimensions, the interference probability remains near the zero-drift Cauchy baseline, reflecting the persistent long-range cross-talk characteristic of diffusion-dominated transport.

% ==============================================================================
% SECTION V: CONCLUSION
% ==============================================================================
\section{Conclusion}\label{sec:conclude}

This letter has characterized the drift-induced transition of FAP distribution from a heavy-tailed Cauchy to an exponentially regularized law. By asymptotically examining the exact FAP density, we identified the CPD, $r_c=\sigma^2/v$, as the fundamental boundary separating diffusion-dominated and drift-dominated regimes. Numerical experiments demonstrated that in low-drift environments, FAP-based communication remains robust, with achievable rates closely following the zero-drift Cauchy baseline. In contrast, variance-matched Gaussian approximations severely underestimate the true communication potential in this regime.

While this mathematical framework provides fundamental insights, it relies on some idealized abstractions as stated in Section~\ref{sec:system}. In practical systems, factors such as bounded parabolic flows \cite{Jamali2019,Farsad2016}, finite-sized receivers governed by binding kinetics \cite{Ahmadzadeh2016}, and particle degradation \cite{Noel2014,Heren2015} will inevitably introduce \textit{secondary} effects that alter large-scale displacement statistics, or physically truncate the heavy tail. 

However, despite these practical complexities, the characteristic scale $r_c$ encapsulates the fundamental physics of the tail transition. Because advection and diffusion are the \textit{primary} drivers of MCvD systems, the spatial crossover from algebraic regime to exponential regularization remains intrinsically anchored at $r_c$. As such, $r_c$ serves as a universal, hardware-independent reference point for assessing feasible receiver spacing and mitigating cross-talk, establishing a principled physical foundation for spatial signaling design for MCvD systems.

\appendices
% ==============================================================================
% APPENDIX: ASYMPTOTIC BEHAVIOR
% ==============================================================================
\section{Asymptotic Behavior of the Modified Bessel Function}
\label{app:bessel}

This appendix summarizes the asymptotic behavior of the modified Bessel function of the second kind, which underlies both the diffusion-dominated core and the drift-regularized tail behaviors analyzed in the main text.

\subsection{Small-Argument Regime \texorpdfstring{($z \to 0^+$)}{(z -> 0+)}}
\label{app:bessel_small}

For the modified Bessel function $K_\nu(z)$ with order $\nu>0$, its small-argument asymptotic expansion is given by~\cite{Olver2010},
\begin{equation}\label{eq:Knu_small}
    K_\nu(z) \sim \frac{1}{2}\Gamma(\nu)\left(\frac{2}{z}\right)^\nu, \qquad z\to 0^+,
\end{equation}
up to lower-order correction terms. In particular, for $\nu=1$, since $\Gamma(1)=1$, the leading-order behavior reduces to
\( 
    K_1(z)\sim \frac{1}{z}.
\)
This algebraic $\mathcal{O}(z^{-1})$ scaling gives rise to the Cauchy-type diffusion-dominated core of the FAP noise distribution discussed in Section~\ref{subsec:regime1}.

\subsection{Large-Argument Regime \texorpdfstring{($z \to \infty$)}{(z -> infinity)}}
\label{app:bessel_large}

For fixed order $\nu$, the large-argument asymptotic expansion of $K_\nu(z)$ is given by~\cite{Olver2010},
\begin{equation}\label{eq:Knu_large}
    K_\nu(z) \sim \sqrt{\frac{\pi}{2z}} \,e^{-z} \left( 1 + \mathcal{O}\!\left(\frac{1}{z}\right) \right), \qquad z\to\infty.
\end{equation}
In particular, for $\nu=1$, the leading-order behavior reduces to
\( 
    K_1(z) \sim \sqrt{\frac{\pi}{2z}} \,e^{-z}.
\)
This exponential decay governs the drift-regularized tail of the FAP noise distribution and leads to the suppression of large transverse displacements described in Section~\ref{subsec:regime2}.

% ==============================================================================
% REFERENCES
% ==============================================================================
\bibliographystyle{IEEEtran}
\bibliography{camera}

\end{document}

%% file: TABLE1.tex
% ==============================================================================
% TABLE OF NOTATIONS
% ==============================================================================
\begin{table}[t]
\centering
\caption{Summary of Main Notations}
\label{tab:notations}
\renewcommand{\arraystretch}{1.15}
\begin{tabular}{|c|p{0.75\linewidth}|}
\hline
\textbf{Notation} & \textbf{Description} \\ \hline
$\mathbf{X}$ 
& Transmitted particle position on the transmitting hyperplane, 
$\mathbf{X}\in\mathbb{R}^{d-1}$. \\ \hline
$\mathbf{Y}$ 
& Received particle position on the transverse (receiving) plane, 
$\mathbf{Y}\in\mathbb{R}^{d-1}$. \\ \hline
$\mathbf{N}$ 
& Random lateral displacement (additive noise) vector of the FAP channel, 
defined as $\mathbf{N}=\mathbf{Y}-\mathbf{X}$. \\ \hline
$\mathbf{n}$ 
& Realization of the noise vector $\mathbf{N}$. \\ \hline
$X$ 
& Scalar transmitted position for the special case $d=2$. \\ \hline
$Y$ 
& Scalar received position for the special case $d=2$. \\ \hline
$N$ 
& Scalar noise random variable for $d=2$, defined as $N=Y-X$. \\ \hline
$n$ 
& Realization of the scalar noise random variable $N$. \\ \hline
$D$ 
& Diffusion coefficient. \\ \hline
$\sigma^2$ 
& Diffusion variance parameter, defined as $\sigma^2=2D$. \\ \hline
$\mathbf{v}$ 
& Constant drift velocity vector, perpendicular to the receiving plane. \\ \hline
$v$ 
& Drift speed, defined as $v=\|\mathbf{v}\|$. (Magnitude of the drift vector)
\\ \hline
$\lambda$ 
& Separation distance between Tx and Rx. \\ \hline
$\|\cdot\|$ 
& Euclidean ($\ell_2$) norm. \\ \hline
$r_c$ 
& The \textit{characteristic propagation distance (CPD)} separating diffusion-dominated and drift-dominated regimes, 
defined as $r_c:=\sigma^2/v$. \\ \hline
$a$ 
& Peak-amplitude constraint on the channel input. (Assuming uniform input $X\sim\mathcal{U}[-a,a]$.) \\ \hline
\end{tabular}
\end{table}